# THEORY AND EXPERIMENT ON THE SURFACE PHOTOVOLTAGE DIFFUSION LENGTH MEASUREMENT IN THIN SILICON WAFERS

**A. BARHDADI**
Laboratoire de Physique des Semiconduteurs et de l'Energie Solaire (P.S.E.S.),
Ecole Normale Supérieure, P.O. Box : 5118, Rabat-10000 (Morocco)
and
The Abdus Salam International Centre for Theoretical Physics, Trieste, Italy.

and

**G. COHEN-SOLAL**
Laboratoire de Physique des Solides et de Cristallogénèse, CNRS-Bellevue,
1, Place Aristide Briand, 92195, Meudon Cedex, France

## Abstract

A simplified theoretical model for the constant-magnitude steady-state surface photovoltage (SPV) technique is developed. Emphasis is placed on the determination of the minority carrier diffusion length L in the particular case for which the sample thickness is quite lower than the true value of L. The model is derived from the theoretical basis with the assumption that incidental light flux varies linearly in the explored range. This model fits very well to experimental data points obtained from both FZ and CZ mono-silicon crystals. We showed that in the case of thin wafers, SPV technique is not adapted to L measurements. It leads only to extracting a short length proportional to the sample thickness. However, if the recombination velocities and the minority carrier diffusion coefficient are known, then we could reach the true value of L by a computer calculation.

# I. INTRODUCTION

The surface photovoltage (SPV) technique is a well-established contact-less experimental tool for the characterisation of semiconductors, which relies on analysing illumination-induced changes in the surface voltage. Recently, a single treatise devoted solely to an in-depth description of this technique has been published [1]. It provides a tutorial review covering nearly all aspects of the technique, from the earliest theoretical and experimental achievements to the latest developments.

One of the most important and successful applications of the SPV technique is the determination of the minority carrier diffusion length L in the bulk. This application has been introduced by GOODMAN [2,3], and well developed by both MOORE [4] and CHIANG et al. [5]. It is certainly the branch of SPV measurements, which has had the most significant impact on the semiconductor manufacturing industry.

Using SPV to determine L parameter has numerous advantages and large success over others techniques [1]. However, it is not frequently employed because of many quantitative limitations of the approach and the validity of assumptions used [1]. One of the necessary experimental conditions required by SPV-based diffusion length measurement states that the sample thickness w must be much larger than L parameter. PHILLIPS [6] has shown that this condition should be met for samples thicker than four times their L. Otherwise, the extracted length is only an effective value, which is not trivially related to any real diffusion length. This effective value should be shorter than the true one because of the recombination of minority charge carriers at the backside of the sample [6]. For much thinner samples, the extracted effective length tends toward a limit value, which should be quite lower than the true value of L. From experiments undertaken by EICHHAMMER [7], this limit value was found to be not much larger than half the sample thickness. But, no explanations related to this particular result have been given. This work is mainly devoted to focus on this point and to try conducting a deep examination of such approaches of the problem. In this aim, we propose a simplified theoretical model and experiment on the relationship between the sample thickness and the extracted minority carrier diffusion length by SPV technique.

## II. MODELLING

Let us consider an n-type semiconductor sample with a w thick and a large front-surface barrier. When monochromatic light, with wavelength $\lambda$ and an incident photon flux density $\Phi_o$, is directed into the surface barrier, the semiconductor absorbs it provided that the photon energy is near or above the band-gap energy of the material. As a result of photon absorption, electron-hole pairs are created. The optical generation rate G(z) of these electron-hole pairs at a distance z from the illuminated surface is [8]:

$$G(z) = \alpha \Phi_o (1-R) \chi \exp(-\alpha z) \qquad (1)$$

where $\alpha$ is the optical absorption coefficient of the semiconductor, R is the optical reflection coefficient at the illuminated surface and $\chi$ is the quantum photon efficiency for electron-hole pair generation. Some of the generated electron-hole pairs recombine immediately at surface edges of the sample. To take this into account, a lower photon quantum efficiency $\eta$ substitutes for $\chi$ in equation (1). Also, while diffusing, a part of the generated charge carriers recombine through complex mechanisms which we can integrate in a simple exponential function as $\exp(-z/L)$ where L is the minority carrier diffusion length in the bulk of the sample.

Based on the above, the hole flux and, then, the photo-courant density $J_p$ resulting from the photo-generated pairs in the whole w thick of the sample may be expressed as:

$$J_p \;=\; C_1 \int_0^w q\,\alpha\,\Phi_o\,(1-R)\,\eta\,e^{-\alpha z}\,e^{-z/L}\,dz \tag{2}$$

where $C_1$ is a constant and q is the absolute value of the electron charge.

In general, $\alpha$, R and $\eta$ all are functions of $\lambda$. Within the wavelength range of interest, however, R and $\eta$ are practically constant. The proportional term $(1-R)\eta$ in (2) will always appear together with $\Phi_o$. For simplicity, we therefore write $\Phi_o(1-R)\eta$ as $\Phi$. So, (2) becomes:

$$J_p \;=\; C_1 \int_0^w q\,\alpha\,\Phi\,e^{-(\alpha+1/L)z}\,dz \tag{3}$$

which, after integration, gives:

$$J_p \;=\; C_1\,q\,\Phi\,\frac{\alpha L}{1+\alpha L}\left[1-e^{-(1+\alpha L)w/L}\right] \tag{4}$$

In the well known first case [2] assuming that the w » L condition is satisfied, the equation (4) is reduced to:

$$J_p \;=\; C_1\,q\,\Phi\,\frac{\alpha L}{1+\alpha L} \tag{5}$$

As the open-circuit photovoltage $V_{oc}$ developed at the illuminated surface barrier is given by [9]:

$$V_{oc} \;=\; \frac{nkT}{q}\,\mathrm{Log}\!\left(\frac{J_p}{J_s}+1\right) \tag{6}$$

then, we can write:

$$J_p \;=\; J_s\left[e^{qV_{oc}/nkT}-1\right] \;=\; C_1\,q\,\Phi\,\frac{\alpha L}{1+\alpha L} \tag{7}$$

where $J_s$ is the saturation current $(J_s \approx q\,D\,p_o/L)$, n is the ideal factor, which is generally close to 1, $p_o$ is the bulk hole density at thermal equilibrium and D is the hole diffusion coefficient.

Keeping $V_{oc}$ at a constant value by adjusting $\Phi_o$ and hence $\Phi$ while varying $\lambda$ and thus $\alpha(\lambda)$, we obtain:

$$\Phi = C_2 \frac{D p_o}{L}\left(\frac{1+\alpha L}{\alpha L}\right) = C_2 \frac{D p_o}{L^2}\left(L + \frac{1}{\alpha}\right) \tag{8}$$

where $C_2$ is also a constant:
$$C_2 = C_1 \left(e^{qV_{oc}/nkT} - 1\right) \tag{8'}$$

This equation is similar to that (expression number 30) formulated by CHIANG [5] when the term $w\alpha$ is very small. From this equation (8), L can be determined by extrapolating the $\Phi$ versus $1/\alpha$ plot to $\Phi = 0$; the negative of the intercept with the $1/\alpha$ axis yields L. Of course, it is assumed that $\alpha(\lambda)$ is known over the required range.

From the same equation (8), we can also easily draw a convenient expression for L as follows:

$$L = \frac{\Phi_1 \alpha_1 - \Phi_2 \alpha_2}{(\Phi_2 - \Phi_1)\alpha_1 \alpha_2} = \frac{\Phi_1 x_2 - \Phi_2 x_1}{(\Phi_2 - \Phi_1)} \qquad \text{where: } x_{1,2} = 1/\alpha_{1,2} \tag{9}$$

Let us consider now the second case for which $w \ll L$. From our knowledge, this particular case has never been well-discussed or/and developed. To consider it, the principle of the method consists in assuming that the flux variation, which maintains the phototension $V_{oc}$ at a constant value, is linear in the exploration zone allowing writing:

$$\Phi\left(\frac{1}{\alpha}\right) = \Phi\left(\frac{1}{\alpha_1}\right) + \left.\frac{d\Phi\left(\frac{1}{\alpha}\right)}{d\left(\frac{1}{\alpha}\right)}\right|_{\frac{1}{\alpha_1}} \left(\frac{1}{\alpha} - \frac{1}{\alpha_1}\right) \tag{10}$$

or:

$$\Phi(x) = \Phi(x_1) + \left.\frac{d\Phi(x)}{dx}\right|_{x_1}(x - x_1) \qquad \text{where } x = 1/\alpha \tag{10'}$$

By extrapolating the $\Phi(x)$ plot to $\Phi = 0$, the negative of the intercept with the x axis yields $\ell$:

$$\ell = \frac{\Phi(x_1)}{\left.\dfrac{d\Phi(x)}{dx}\right|_{x_1}} - x_1 \tag{11}$$

with $\Phi(x)$ can be expressed from the equation (4) as follows:

$$\Phi(x) = \frac{J_p}{q} \frac{x+L}{L} \frac{1}{1 - e^{-\left(\frac{w}{L} + \frac{w}{x}\right)}} \tag{12}$$

It should be noted that the first term of the second member in the equation (11) is, in fact, the inverse of the logarithmic derivative of $\Phi(x)$ at $x_1$:

$$\frac{\Phi(x_1)}{\left.\dfrac{d\Phi(x)}{dx}\right|_{x_1}} = \frac{1}{\left.\dfrac{d}{dx} \text{Log}\,\Phi(x)\right|_{x_1}} = \left( \frac{1}{x_1 + L} + \frac{w\, e^{-\left(\frac{w}{L} + \frac{w}{x_1}\right)}}{x_1^2 \left[1 - e^{-\left(\frac{w}{L} + \frac{w}{x_1}\right)}\right]} \right)^{-1} \tag{13}$$

When we put this in the equation (5), it becomes after deduction:

$$\ell = \frac{x_1^2 L - (x_1 w + w L + x_1 L) x_1 e^{-\left(\frac{w}{L} + \frac{w}{x_1}\right)}}{x_1^2 + (x_1 w + w L - x_1^2) e^{-\left(\frac{w}{L} + \frac{w}{x_1}\right)}} \tag{14}$$

Let us perform now a development in series of the exponential function up to the second order as follows:

$$e^{-\left(\frac{w}{L} + \frac{w}{x_1}\right)} \cong 1 - \left(\frac{w}{L} + \frac{w}{x_1}\right) + \frac{1}{2}\left(\frac{w}{L} + \frac{w}{x_1}\right)^2 \tag{15}$$

When we put this in the equation (14), we accomplish a simple but a little long and tiresome process to lead to:

$$-\ell \cong \frac{1}{2} w \tag{16}$$

This result, remarkable by its simplicity, means that in any case where the sample thickness is quite smaller than the minority charge carrier diffusion length L, the measured lengths $\ell$ using SPV technique are nothing of the kind. They only give an adequate value of the sample thickness.

It could be interesting to mention that CHIANG formulas [5] are less different to those we are proposing in this work. This difference should be due to the more explicit details given by this author on the minority carrier losses by recombination processes. Indeed, from our side, these losses have been entirely integrated in a simplified expression which leads not only to the same conclusions on the determination of L, but also specify the sense to give to the SPV measurements when using thin wafers.

## III. EXPERIMENTAL VALIDATION

To valid the remarkable theoretical result expressed by equation (16), some experimental measurements have been performed in the required corresponding conditions. They have been conducted on virgin, n-type, phosphorus doped, <100>, 1-5 $\Omega$cm mono-silicon crystals from Wacker. Both Czochralski pulled (CZ) and float-zone grown (FZ) single crystals have been used. They present a very similar initial true diffusion length L with a predetermined average value around 500 µm. Samples of dimensions typically 1x1 cm$^2$ area were cut out from circular wafers of 3 in. diameter and 450 µm thick. These samples were first lapped on both faces with Aloxite 125 grit (average grain size ≈ 9 µm) to perform serial deeper and deeper sectioning allowing to obtain a series of thickness w ranging from 150 to nearly 400 µm in variable steps as measured with a micrometer. Next, one face (i.e., the front side) of each sample was slightly etched in (5 : 3 : 3) 70% HNO$_3$ : 50% HF : 100% CH$_3$COOH chemical mixture for 60 s, which would remove roughly 15 µm thick. After, all prepared samples were degreased in trichlorethylene, cleaned in acetone, carefully rinsed in running de-ionised water and finally dried under a nitrogen gas flow to be ready for the SPV measurements in the particular w « L experimental condition leading to extract $\ell$ effective values instead of L.

$\ell$ was measured on the etched face of the sample by means of a home built SPV set-up using a small pickup probe (Ø = 1 mm). To have a good SPV signal, all samples were boiled in de-ionised water for 1 hour to ensure correct band bending at their front surface. The relative error in extracted $\ell$ value is estimated at 10 %. Just before the measurements, the samples were cleaned again to minimise external contamination and remove the native oxide layer.

Figures 1 and 2 show the dependence of extracted $\ell$ on the sample thickness in the case of FZ and CZ crystals respectively. Each $\ell$ value reported in these figures has been obtained from repeated measure performed on nearly the same area of the sample. The straight line fit to FZ data points and that of CZ, show very similar slopes. From these experimental results, it is very easy to point out that, at least in the chosen thickness range, the measured length $\ell$ varies quite proportionately with the sample thickness w according to that expected and predicted by the theoretical model described above. As shown, we find slopes around an average value of 0,5 for the two kinds of crystals. This result agrees very well with that explain-less empirically deduced from the experiments undertaken by EICHHAMMER [7].

Using the whole equation (14), when we plot the evolution of $\ell$ as function of the sample thickness w (figure 3), one can show clearly that for some L, w and $x_1$ values, corresponding roughly to our experimental conditions, the linear approximation used in the model is well justified. We find then slopes of 0,5 average value, with small variations to be associated, in less explicit way, with the recombination mechanisms and the true diffusion lengths. As we can see in the figure 3, within the framework of our assumption, we should have to use samples with thickness at least three times larger than the diffusion length to hope performing a correct measurement of this parameter by SPV technique.

Although the curve slope depends partly on the diffusion length value, we saw that its variations remain small and could not lead to any determination of L except to determine, in other respects, parameters such as D, $p_o$ as well as recombination velocities in the bulk and at the involved surfaces of the sample.

## IV. CONCLUSION

In this work, another dimension in the influence of the wafer thickness on the minority carrier diffusion length L measurement by SPV technique emerges. It is shown that in the case of wafer with low thickness compared to L, SPV technique was not adapted to L measurements since it amounts to extract a length proportional to the sample thickness. If, by another way, the values of the recombination velocities in the space charge region as well as on the involved surfaces of the sample and D coefficient are known, then L should be obtained by using computer processing either by CHIANG model [5] or from equation 14. This requires first measuring the optical reflection and absorption coefficients, the doping density in the sample, the open-circuit photovoltage at the illuminated surface and the incident photon flux density. Moreover, it needs to make simple assumptions on the low-level excitation and on the linearity of Φ function.

# ACKNOWLEDGMENTS

This work was achieved during the visit of the first author as a regular associate scientist at the Abdus Salam International Centre for Theoretical Physics (ICTP), Trieste. This author would like to thank the director and staff of this centre for the generous hospitality, the great support and the efficient assistance. He also wishes to thank Dr. J-C. Muller from PHASE Laboratory, CNRS-Strasbourg, France, for his precious help and scientific assistance during the experimental study. Many thanks also to Prof. G. Furlan for carefully reading the manuscript and for his kind collaboration.

# REFERENCES

[1]    L. KRONIK and Y. SHAPIRA
       Surface Science Reports, 37 (1-5), 1999, pp. 1-206.

[2]    A. M. GOODMAN
       J. Appl. Phys., 32 (12), 1961, p. 2550.

[3]    Annual Book of ASTM Standards
       ASTM, Philadelphia, 1982.

[4]    A. R. MOORE
       J. Appl. Phys., 54 (1), 1983, p. 222.

[5]    C-L. CHIANG and S. WAGNER
       IEEE Trans. on Elec. Devices, ED-32 (9), 1985, p. 1722

[6]    W. E. PHILLIPS
       Sol. State Electr., 15, 1972, p. 1097.

[7]    W. EICHHAMMER, Vu-Thuong-QUAT and P. SIFFERT
       J. Appl. Phys., 66 (8), 1989, p. 3857

[8]    S. M. SZE
       Physics of Semiconductor Devices, $2^{nd}$ Ed., New York, Wiley, 1981

[9]    C. Y. CHANG and S. M. SZE
       Sol. State Electr., 13 (1970) p. 727

# FIGURE CAPTIONS

*Figure 1:*

Dependence of the SPV extracted $\ell$ value on the FZ mono-silicon crystal wafer thickness. The straight line fit to data points shows a slope of about 0,57.

*Figure 2:*

Dependence of the SPV extracted $\ell$ value on the CZ mono-silicon crystal wafer thickness. The straight line fit to data points shows a slope of about 0,55.

*Figure 3:*

Dependence of the calculated $\ell$ value on the semiconductor wafer thickness. For some L, w and $x_1$ values, corresponding roughly to our experimental conditions, the linear approximation used in the model is well justified. We find then slopes of 0,5 average value, We should have to use samples with thickness at least three times larger than the diffusion length to hope performing a correct L measurement by SPV.

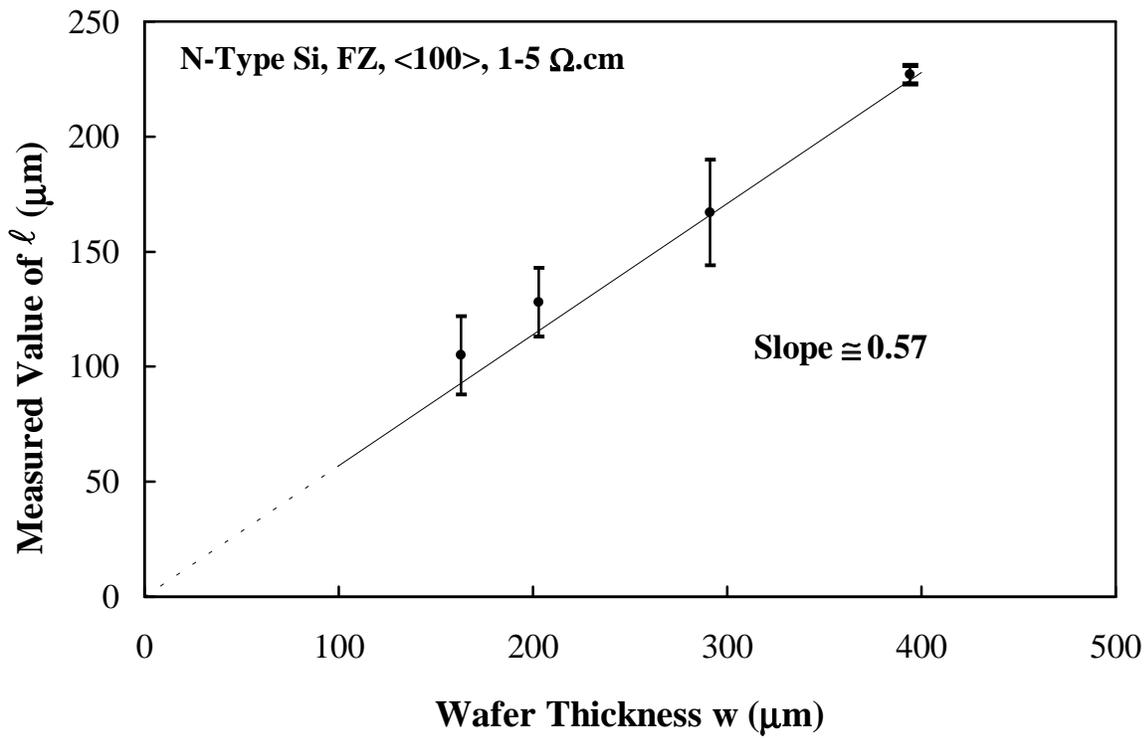

*FIGURE 1*

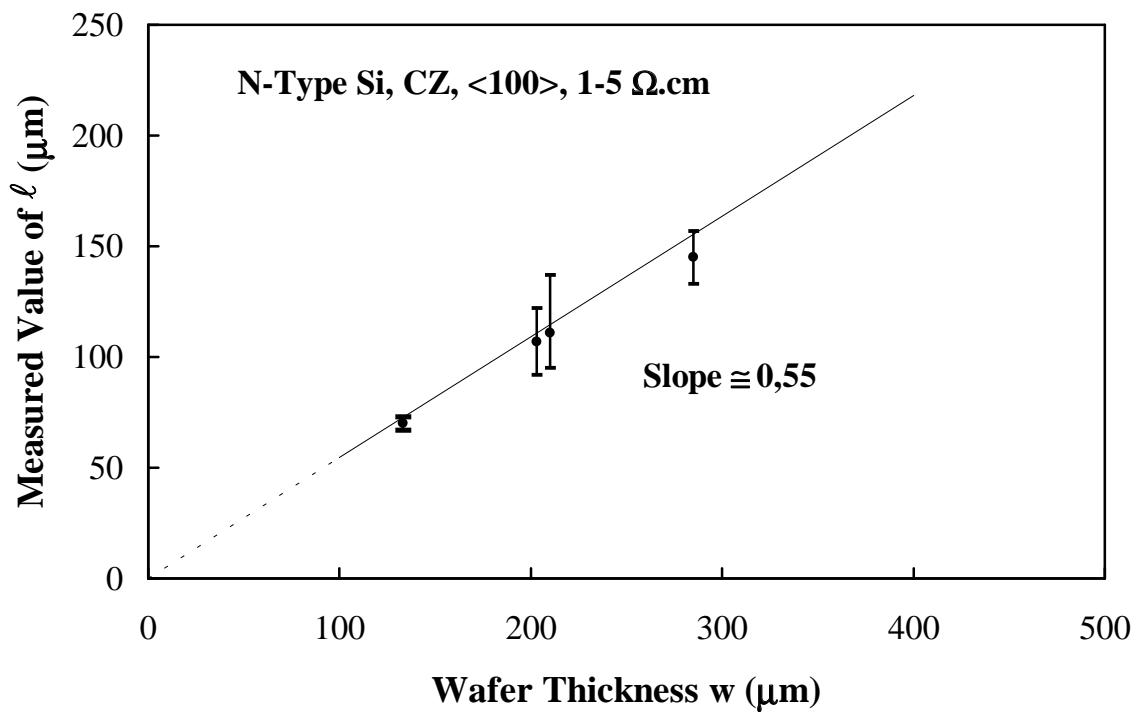

*FIGURE 2*

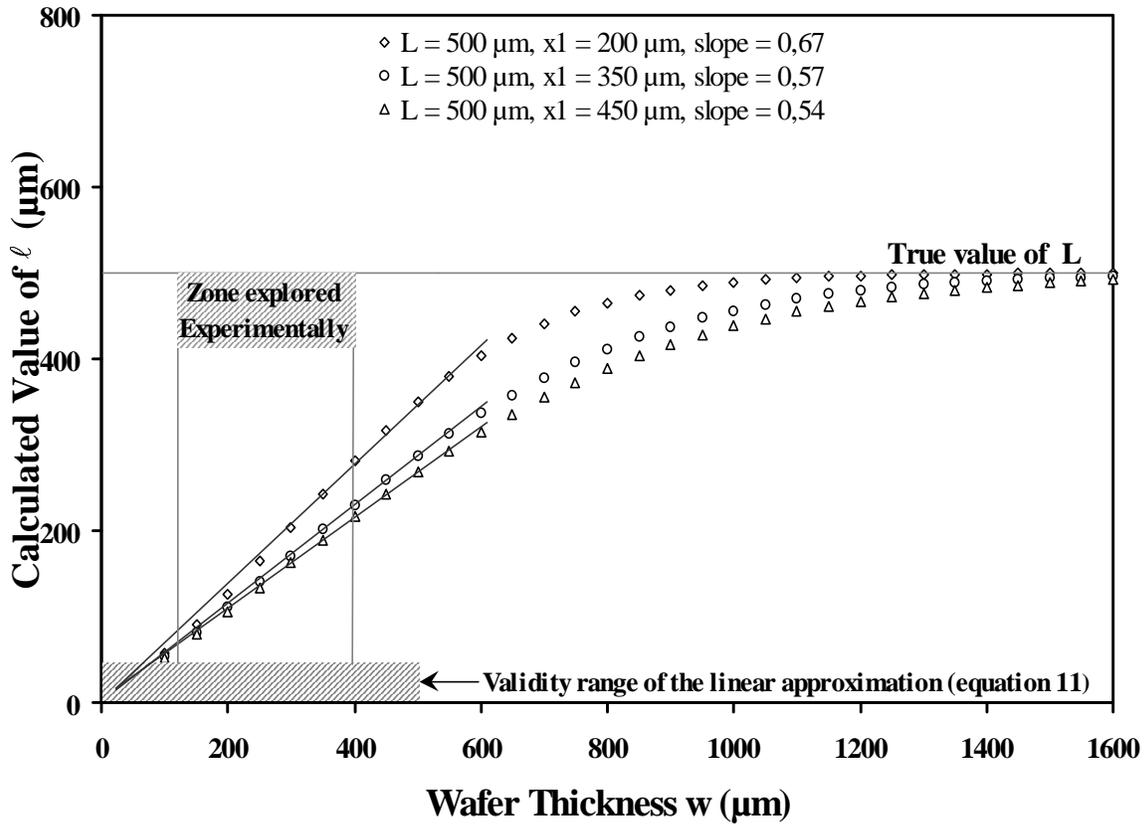

*FIGURE 3*